# Room Temperature de Haas – van Alphen Effect in Silicon Nanosandwiches

N.T. Bagraev[1,2], V.Yu. Grigoryev[2], L.E. Klyachkin[1], A.M. Malyarenko[1], V.A. Mashkov[2] and V.V. Romanov[2]

[1]Ioffe Physical-Technical Institute, 194021, St. Petersburg, Russia

[2]Peter the Great St. Petersburg Polytechnic University, 195251, St. Petersburg, Russia

e-mail: bagraev@mail.ioffe.ru

## Abstract

The negative-$U$ impurity stripes confining the edge channels of semiconductor quantum wells are shown to allow the effective cooling inside in the process of the spin-dependent transport. The aforesaid promotes also the creation of composite bosons and fermions by the capture of single magnetic flux quanta on the edge channels under the conditions of low sheet density of carriers, thus opening new opportunities for the registration of the quantum kinetic phenomena in weak magnetic fields at high temperatures up to the room temperature. As a certain version noted above we present the first findings of the high temperature de Haas-van Alphen, 300K, and quantum Hall, 77K, effects in the silicon sandwich structure that represents the ultra-narrow, 2 nm, $p$-type quantum well (Si-QW) confined by the delta barriers heavily doped with boron on the $n$-type Si (100) surface. These data appear to result from the low density of single holes that are of small effective mass in the edge channels of $p$-type Si-QW because of the impurity confinement by the stripes consisting of the negative-$U$ dipole boron centers which seems to give rise to the efficiency reduction of the electron-electron interaction.

The Shubnikov – de Haas (ShdH) and de Haas – van Alphen (dHvA) effects as well as the quantum Hall effect (QHE) are quantum phenomena that manifest themselves at the macroscopic level and have attracted a lot of attention, because they reveal a deeper insight into the processes due to charge and spin correlations in low-dimensional systems[1,2]. Until recently the observation of these quantum effects in the device structures required ultra-low temperatures and ultra-high magnetic fields[3]. Otherwise there were the difficulties to provide small effective mass, $m^*$, and long momentum relaxation time, $\tau_m$, of charge carriers, which result from the so-called strong field assumption, $\mu B \gg 1$, where $\mu = e\tau_m/m^*$ is the mobility of the charge carriers. This severe criterion along with the condition $\hbar\omega_c \gg kT$ hindered the application of the ShdH-dHvA-QHE techniques to control the characteristics of the device structures in the interval between the liquid-nitrogen and room temperatures, where $\hbar\omega_c$ is the energy gap between adjacent Landau levels, $\omega_c = eB/m^*$ is the cyclotron frequency. Nevertheless, the ShdH oscillations were observed at room temperature in graphene, a single layer of carbon atoms tightly packed in a honeycomb crystal lattice, owing to the small effective mass of charge carriers, $\sim 10^{-4} m_0$, although the magnetic field as high as 29T was necessary to be used because of relatively short momentum relaxation time[4-6]. Thus, the problem of the fulfillment of the strong field assumption at low magnetic fields has remained virtually unresolved. Perhaps, its certain decision is to use the pairs of edge channels in topological two-dimensional insulators and superconductors, in which the carriers with anti-parallel spins move in opposite directions[7]. Especially as recently the ideas are suggested that the mobility and spin-lattice relaxation time of the carriers in topological channels can be increased, if to hide them in the cover consisting of the $d$- and $f$-like impurity centers[8,9]. Here we use as similar clothes the striations of the negative-$U$ dipole boron centers that allow except noted advantages to achieve the effective cooling inside the edge channels of semiconductor quantum wells in the process of the spin-dependent transport.

**The aforesaid promotes also the creation of composite bosons and fermions by the capture of single magnetic flux quanta on the edge channels under the conditions of low sheet density of carriers, thus opening new opportunities for the registration of the quantum kinetic phenomena in weak magnetic fields at high temperatures up to the room temperature. As a certain version noted above we present the first findings of the high temperature de Haas-van Alphen, 300K, and quantum Hall, 77K, effects in the silicon sandwich structure that represents the ultra-narrow, 2 nm, *p*-type quantum well (Si-QW) confined by the delta barriers heavily doped with boron on the *n*-type Si (100) surface. These data appear to result from the low density of single holes that are of small effective mass in the edge channels of *p*-type Si-QW because of the impurity confinement by the stripes consisting of the negative-*U* dipole boron centers which seems to give rise to the efficiency reduction of the electron-electron interaction.**

The device was prepared using silicon planar technology. After precise oxidation of the n-type Si (100) wafer, making a mask and performing photolithography, we have applied short-time low temperature diffusion of boron from a gas phase[10,11]. Finally, the ultra-shallow $p^+$-$n$ junction has been identified, with the $p^+$ diffusion profile depth of 8 nm and the extremely high concentration of boron, $5 \cdot 10^{21}$ cm$^{-3}$, according to the SIMS data (Fig. 1a)[10]. Next step was to apply the cyclotron resonance (Fig. 1b), the electron spin resonance, the tunneling spectroscopy (Fig. 1c), infrared Fourier spectroscopy methods as well as the measurements of the quantum conductance staircase for the studies of the quantum properties of the silicon nanosandwich structures[11-14].

Firstly, the cyclotron resonance angular dependences have shown that the $p^+$ diffusion profile contains the ultra-narrow p-type silicon quantum well, Si-QW, confined by the wide-gap delta-barriers heavily doped with boron (Fig. 1a)[12,13]. Secondly, the one-electron band scheme for the delta barriers and the energy positions of two-dimensional subbands of holes in the Si-QW have been revealed using the tunneling and the infrared Fourier spectroscopy techniques (Fig. 1d)[11,13].

Then, the studies of the spin interference by measuring the Aharonov – Casher oscillations allowed the identification of the extremely low value of the effective mass of holes[11]. These data have been also confirmed by measuring the temperature dependence of the ShdH and dHvA oscillations[13,15].

The planar silicon sandwich structures prepared were very surprised to demonstrate the high mobility of holes in the Si-QW in spite of the extremely high concentration of boron inside the delta barriers. Specifically, the cyclotron resonance spectra exhibit the long moment relaxation time for both heavy and light holes, >5 $10^{-10}$ s, and electrons, >2 $10^{-10}$ s (Fig. 1b)[12,13]. These results appeared to be caused by the formation of the trigonal dipole boron centers, $B^+$ - $B^-$, due to the negative-$U$ reaction: $2B^o \rightarrow B^+ + B^-$ (Fig. 1e)[11,14]. The excited triplet states of the negative-$U$ centers were observed firstly by measuring the electron spin resonance angular dependences. It is important that the ESR spectra of the negative-$U$ centers are revealed only after cooling in magnetic fields above critical value of 0.22T, with persistent behavior in the dependence of temperature variations and crystallographic orientations of magnetic field under cooling procedure[16]. Such persistent behavior of the ESR spectra seems to be evidence of the dynamic magnetic moment due to the arrays of the trigonal dipole boron centers which dominate inside the delta-barriers confining the $p$-type Si-QW[11,13,16].

The scanning tunneling microscopy, STM, as well as the scanning tunneling spectroscopy, STS, studies showed that the negative -$U$ dipole boron centers are able to form the stripes crystallographically oriented along the [110] axes (Fig. 1f and 1g)[11]. Moreover, the subsequences of these stripes are able to create the edge channels that define the conductance of the silicon nanosandwiches. And, if the sheet density of holes is rather small, on each such stripe no more than one hole settles down that appears to result in the neutralization of the electron-electron interaction owing to the exchange interaction with the negative-$U$ dipole boron centers. Therefore we could apparently control the step-by-step capture of single magnetic flux quanta on the negative-$U$ dipole boron stripes containing single holes by measuring the field dependences of the static magnetic susceptibility. We have carried out these experiments with the long edge channel, 2 mm, of the *p*-type Si-QW, taking account of the sheet density of holes, $3\times10^{13}$ m$^{-2}$, measured with similar device prepared within frameworks of the Hall geometry (Fig. 1h). This value of the sheet density of holes specified previously that each stripe seems to contain a single hole in this edge channel (Fig. 1i).

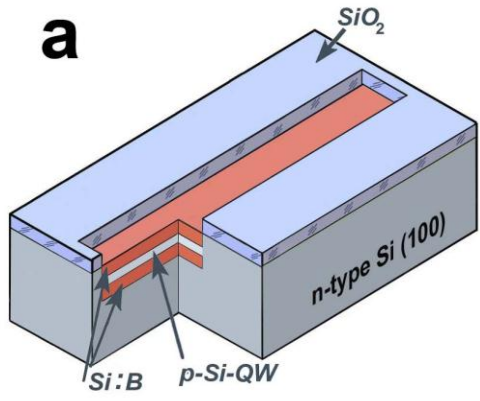
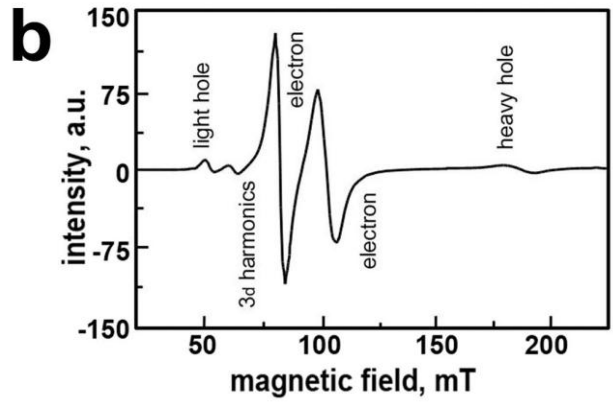
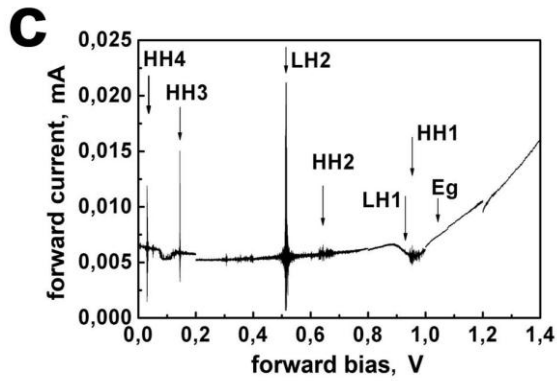
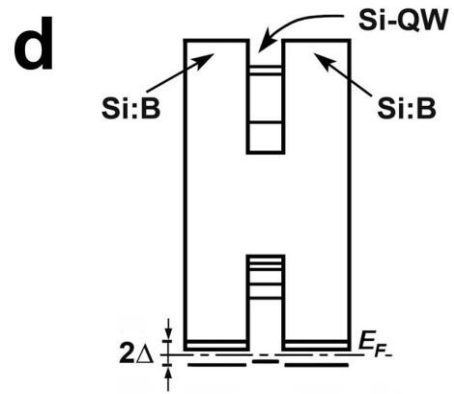
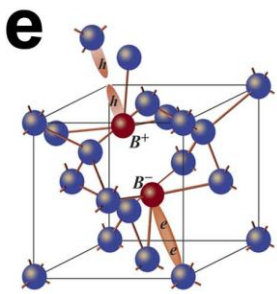
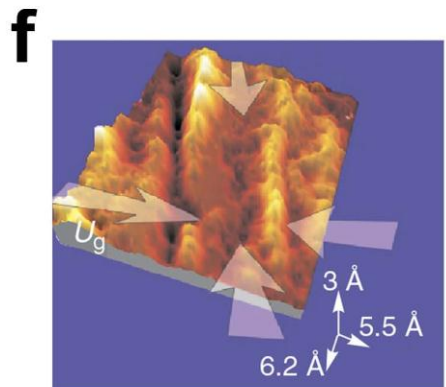
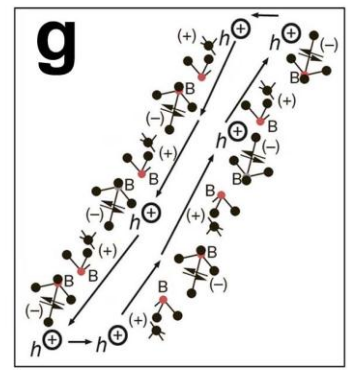
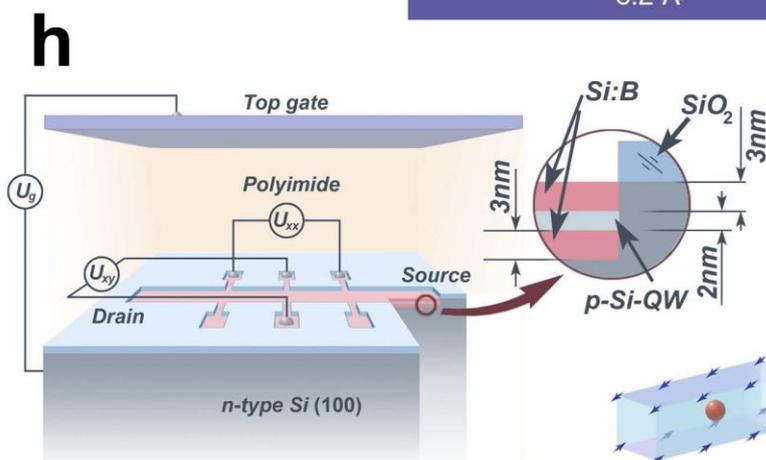
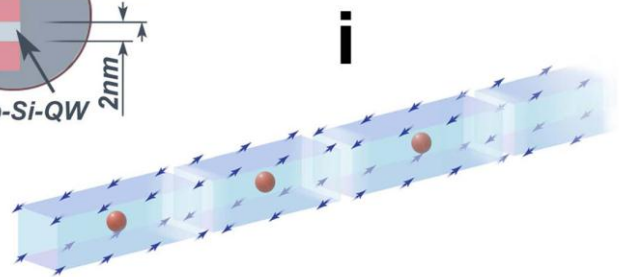

**Figure 1. Planar silicon nanosandwiches.**
(a) Scheme of the silicon nanosandwich that represents the ultra-narrow *p*-type silicon quantum well (Si-QW) confined by the delta barriers heavily doped with boron on the *n*-type Si (100) wafer.
(b) Cyclotron resonance (CR) spectrum for a QW $p^+$-*n* junction on {100}-silicon (500Ohm·cm), magnetic field B within the {110}-plane perpendicular to the {100}-interface (B∥<100> +30°); *T*=4.0K, *v*=9.45 GHz.
(c) The current-voltage characteristics under forward bias applied to the silicon nanosandwich. The energy position of each subband of 2D holes in Si-QW is revealed as a current peak under optimal tunneling conditions when it coincides with Fermi level. *T*=300K.
(d) The one-electron band scheme of the *p*-type Si-QW confined by the δ-barriers heavily doped with boron on the *n*-type Si (100) surface. 2Δ depicts the correlation gap in the delta barriers that results from the formation of the negative-*U* dipole boron centers.
(e) Model for the elastic reconstruction of a shallow boron acceptor which is accompanied by the formation of the trigonal dipole ($B^+$ - $B^-$) centers as a result of the negative-*U* reaction: $2B^0 \rightarrow B^+ + B^-$.
(f) STM image of the upper delta barrier heavily doped with boron that demonstrates the chains of dipole boron centers oriented along the [011] axis.
(g) The fragments of the edge channels that contain the stripes consisting of the negative U dipole boron centers.
(h) Experimental device prepared within the Hall geometry based on an ultra-narrow *p*-type silicon quantum well (Si-QW) confined by the delta barriers heavily doped with boron on the *n*-Si (100) surface.
(i) The model of the edge channel in the silicon nanosandwich that is confined by the impurity stripes containing single holes.

Figures 2a and b show the room temperature field dependences of the static magnetic susceptibility of the silicon nanosandwiches that have been obtained by the Faraday method. The high sensitivity, $10^{-9}$–$10^{-10}$ GGS, of the balance spectrometer MGD31FG provided the high stability that is necessary to calibrate the *BdB/dx* values. In turn, for the *BdB/dx* calibration, we used pure InP single crystals that are similar to investigated samples in shape and size, and reveal the high temperature stability of the magnetic susceptibility value, $\chi = -313 \times 10^{-9}$ cm$^3$/g[13,15].

Owing to so high sensitivity of the balance spectrometer up to weak magnetic fields, the oscillations of the static magnetic susceptibility that is due to the step-by-step capture of single magnetic flux quanta appeared to be revealed against the background of the strong diamagnetism of the negative-*U* dipole boron stripes (Figs. 2c and d). The half-circle value of these oscillations is evidence of such an assumption, with the alternation of the spin-up and spin-down for holes under these step-by-step processes in the edge channel, if the size of the edge channel is taken into account, $\Phi_0 = \Delta\Phi = \Delta B \cdot S$, where $\Delta B = 10G$; $S = 2$ mm x 2 nm $= 4\cdot10^{-12}m^2$ is the area of the edge channel; $\Phi_0=h/e$ is the single magnetic flux quantum. Thus, the step-by-step change of a magnetic field that is equal to $\Delta B = 10G$ is able to result in the capture of single magnetic flux quanta on each hole after 124 circles, which correspond approximately to the number of them in the edge channel in view of the sheet density value. Under these conditions, $\Phi_0 = \Delta B \cdot S = B_0 \cdot S_0$, where $B_0 = 124 \cdot \Delta B = 1240G$, $S_0$ is the area occupied by a single hole inside the negative-$U$ dipole boron stripes which is defined by the middle distance between holes in the edge channel, $S_0 = S/124 = 16.6$ microns. So, there is a unique possibility to create composite bosons since weak magnetic fields that result from the capture of single magnetic flux quanta on the holes which are poorly interacting with each other in the presence of the negative-$U$ dipole boron stripes.

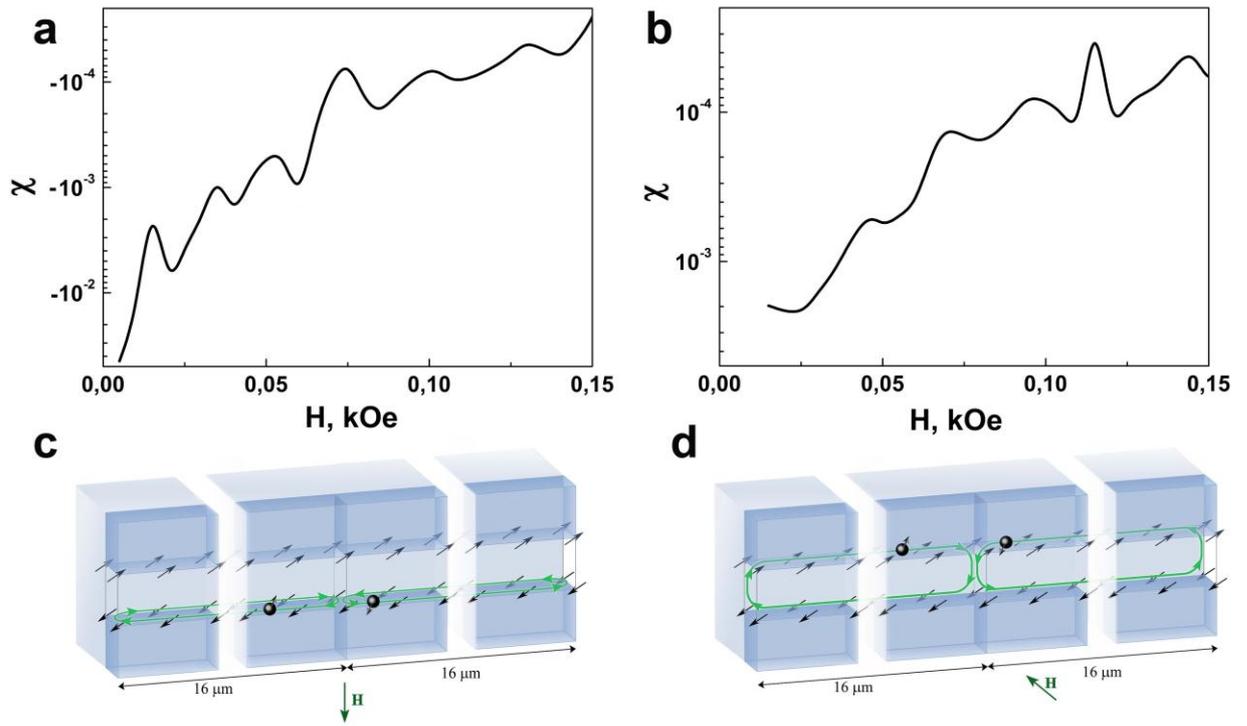

**Figure 2. Step-by-step capture of magnetic flux quanta on the edge channel in silicon nanosandwich.**
(a) and (b) The field dependences of the static magnetic susceptibility at room temperature in silicon nanosandwich, $p_{2D}=3\times10^{13}m^{-2}$, which reveal strong diamagnetism that is caused by the spin precession of single holes confined by the negative-$U$ dipole boron stripes in the edge channel and the oscillations due to the step-by-step capture of single magnetic flux quantum on the edge channel at perpendicular (a and c) and parallel (b and d) orientation of a magnetic field related to the Si-QW plane. $T=300K$.

These reasons are supported by the analysis of huge amplitudes of the static magnetic susceptibility oscillations which appears to give rise to the values of magnetization, $J=\chi H$, up to 0.2 Oe. From here it is possible to estimate the value of the magnetic moment created by the capture of the magnetic flux quantum on a single hole in view its volume, $M=J \cdot V_0$, where $V_0 = L \cdot S_1$; $L$ is the middle distance between holes in the edge channel and $S_1$ is the cross section of the edge channel. This assessment results in the value of the magnetic moment, $2.4 \cdot 10^4$ $m_B$, that unambiguously points to a fundamental role of the stripes consisting of the negative $U$ dipole boron centers in its formation, where $m_B$ is the Bohr magneton. And really, if we estimate the number of these centers in the volume occupied by a single hole in view of the concentration of boron, $5 \cdot 10^{21}$ cm$^{-3}$, and having compared everyone the magnetic moment equal to the Bohr magneton, it is possible to be convinced that it practically coincides with the value given above. Similar results are important to be obtained in the case of both parallel and perpendicular orientation of the magnetic field to the Si-QW plane, because the edge channel has square section (see Figs. 2c and d). It is also necessary to note supervision of almost limit value of a diamagnetic static susceptibility in weak magnetic fields, $\chi = 1/4\pi$, that appears to be due to superconducting properties of edge channels because of the formation of the negative-$U$ dipole boron centers.

Thus, the magnetic susceptibility response to the capture of magnetic flux quanta on the edge channel is caused by the magnetic ordering of the stripes through single holes. This exchange interaction seems to lead to partial localization of single holes and as a result to the reduction of electron-electron interaction. Moreover, by increasing the magnetic field the static magnetic susceptibility begins to reveal the dHvA oscillations due to the creation of the Landau levels, $E_v = \hbar\omega_c(v + \frac{1}{2})$, $v$ is the number of the Landau level. In particular, a prerequisite to cover the edge channel by single magnetic flux quanta appears to be accomplished when the external magnetic field is equal to $B_0 = 1240$G, see the relationship presented above, $\Phi_0 = \Delta B \cdot S = B_0 \cdot S_0$, that is consistent with the first Landau level feeling, $v_1 = 1$, $v_1 = p_{2D} \cdot h/e\, B_0$.

We studied carefully the dHvA oscillations created at both parallel and perpendicular orientation of the magnetic field to the Si-QW plane and registered except the dips related to the Landau levels, $v = 1, 2, 3, 4, 5, 6$, the fractional peaks, $v = 4/3$ and $v = 5/3$ (Figs. 3a-f). These findings demonstrate that in certain ratio between the number of magnetic flux quanta and single holes, $1/v$, both composite bosons and fermions seem to be proceeded by step-by-step change of a magnetic field. Besides, using specific diagram of magnetic field – edge channel feeling, the variations of integer and fractional values of $v$ illustrate conveniently their relationship by varying the external magnetic field (Fig. 4). It should be noted also that the strong diamagnetism of the negative $U$ impurity stripes surrounding the Si-QW edge channels allowed the observation of the room temperature hysteresis of the dHvA oscillations dependent on the proximity of the Landau and Fermi levels. In turn, the creation of the composite bosons ever in weak magnetic fields inside stripes containing single holes can lead to the emergence of the Faraday effect under the conditions of the source - drain current in the edge channel, $I_{ds} = dE/d\Phi$. This model has been suggested by Laughlin to account for the quantum staircase of the Hall resistance[17]. Here we present the results of the measurements of not only integer but also fractional quantum Hall effect in the same magnetic field as well as the dHvA oscillations, $I_{ds} = I_{xx} = eU_{xy}/(1/v)\Phi_0$, $=> G_{xy} = ve^2/h$, where $v$ can accept both integer and fractional values.

Firstly, the SdH oscillations and the quantum Hall staircase are demonstrated, with the identification of both the integral and fractional quantum Hall effects (Figs. 5a, b and c). Secondly, the range of a magnetic field corresponding to the Hall plateaus and the longitudinal "zero" resistance is in a good agreement with the interval of the dHvA oscillations thereby verifying the principal role of the Faraday effect in these processes. Here, the confinement of single holes inside the stripes consisting of the negative-$U$ dipole boron centers seems to lead not only to the efficiency reduction of the electron-electron interaction, but also promotes quantization of the interelectronic spacing[18-21]. Thus, the stabilization of a ratio between the number of magnetic flux quanta and single holes in edge channels, $1/\nu$, is reached at certain values of an external magnetic field, thereby promoting the registration of both integer, and fractional quantum Hall effect[22,23].

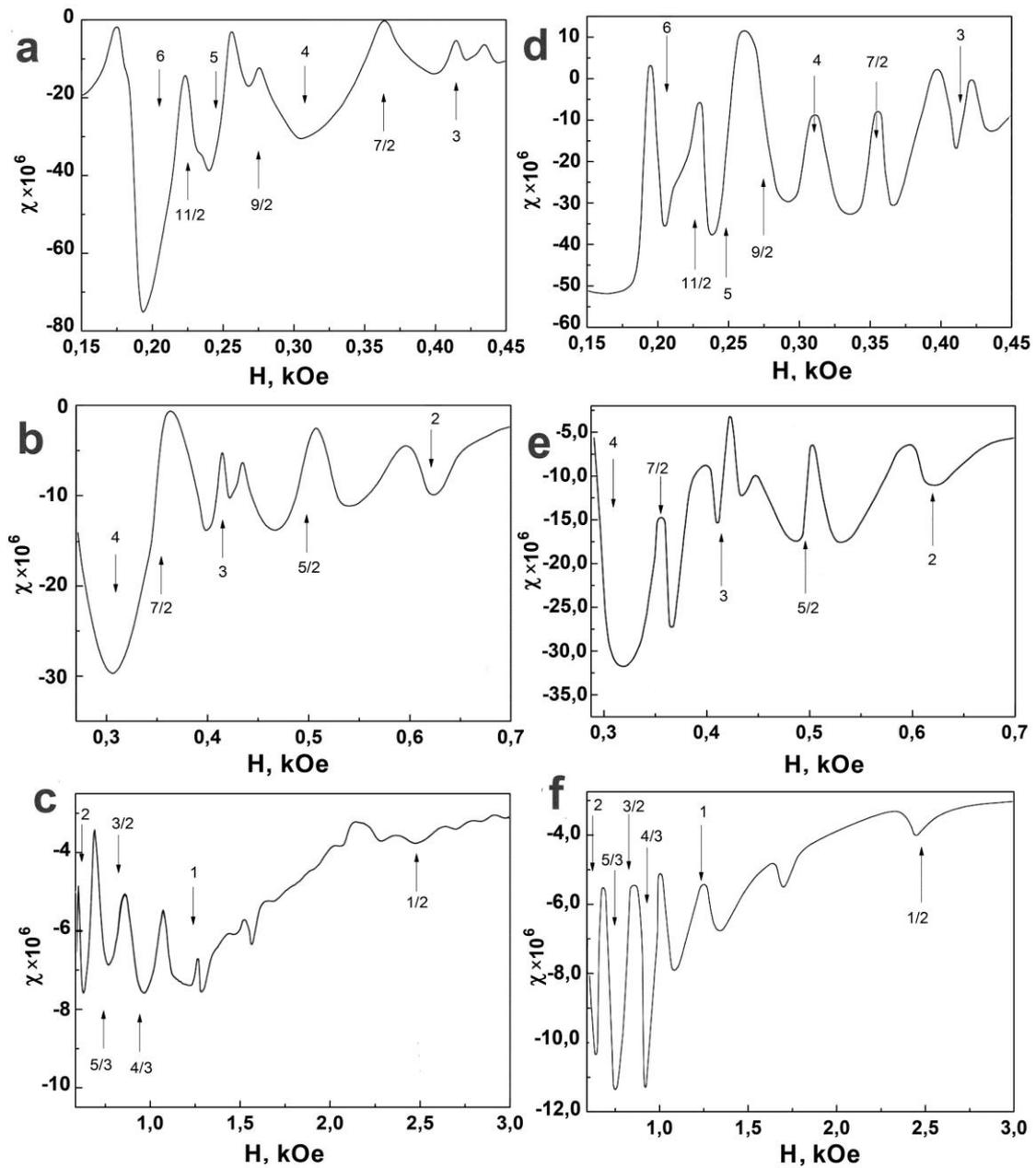

**Figure 3. The de Haas – van Alphen oscillations.**
The de Haas – van Alphen oscillations revealed in the field dependences of the static magnetic susceptibility measured at room temperature in silicon sandwich, $p_{2D}=3 \times 10^{13} m^{-2}$, at perpendicular (a, b, c) and parallel (d, e, f) orientation of a magnetic field related to the Si-QW plane. The dips is related to the Landau levels $\nu=1, 2, 3, 4, 5, 6$. $T=300K$.

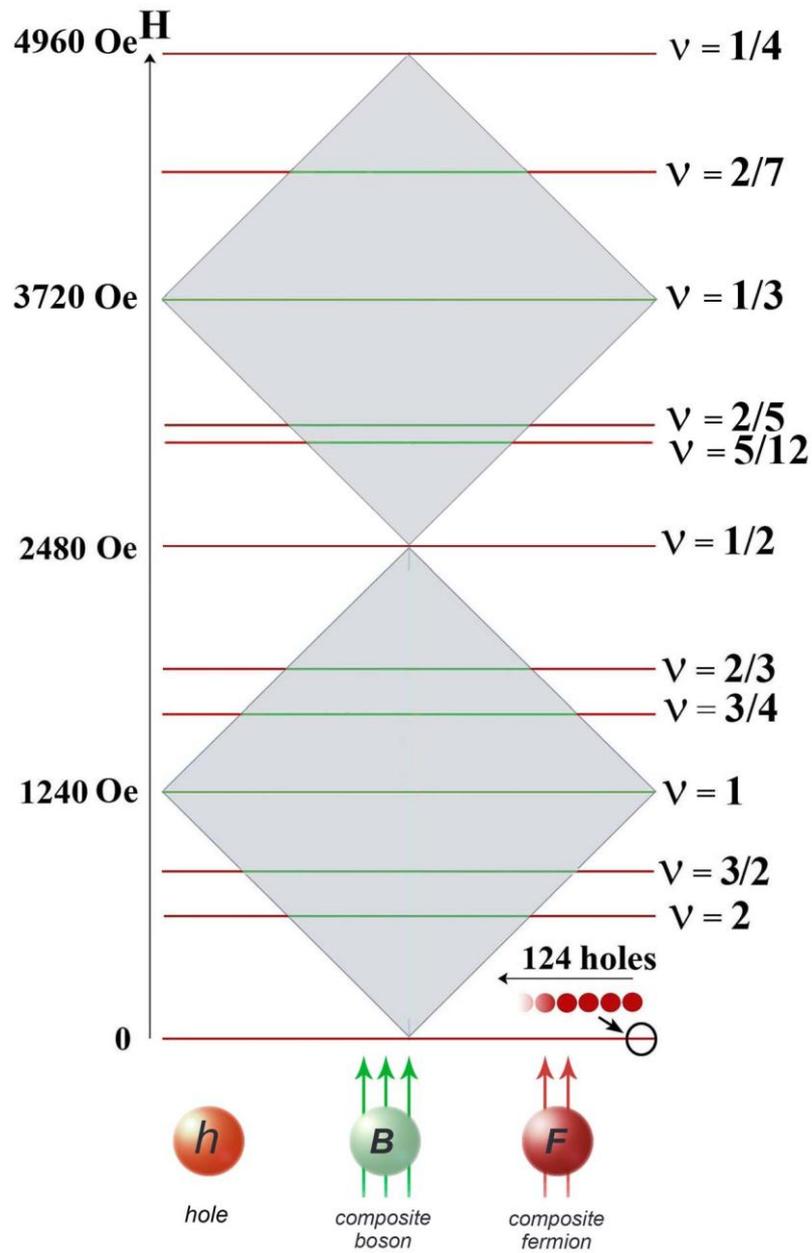

**Figure 4. Composite bosons and fermions.**
Diagram showing the subsequent creation of composite bosons (green line) and fermions (red line) that result from the capture of single magnetic flux quantum at the edge channel of silicon nanosandwich confined by the negative-*U* impurity stripes containing single holes.

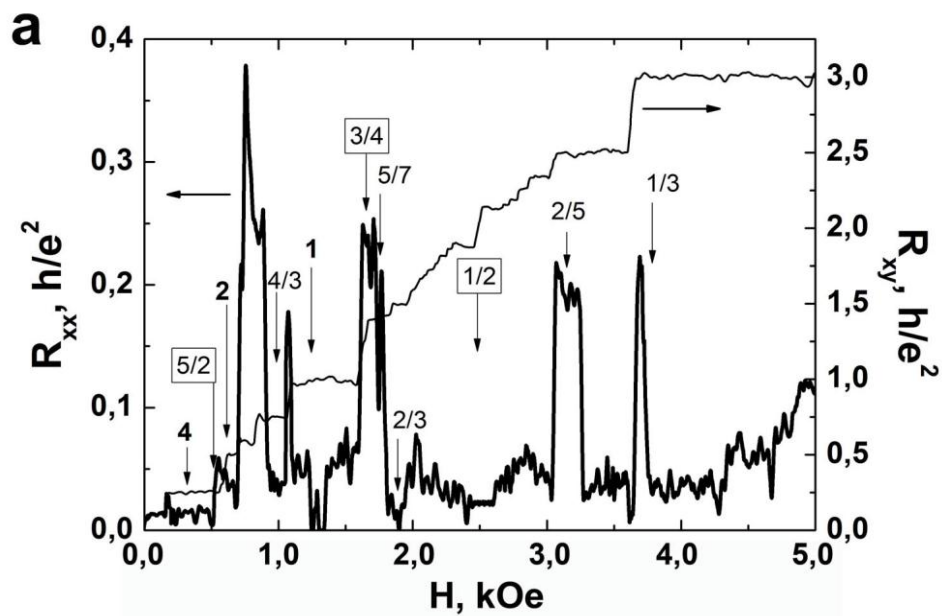
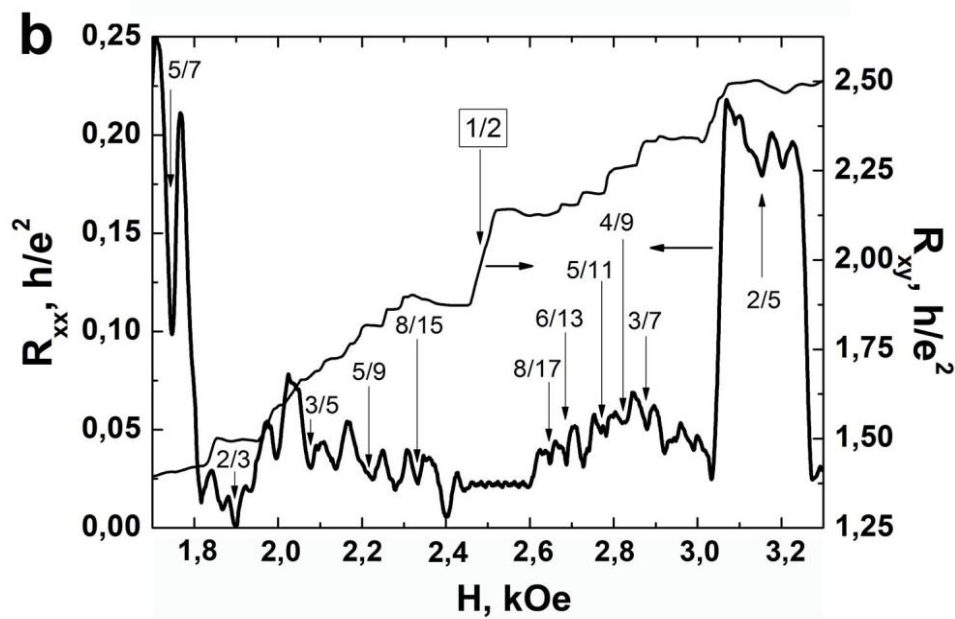
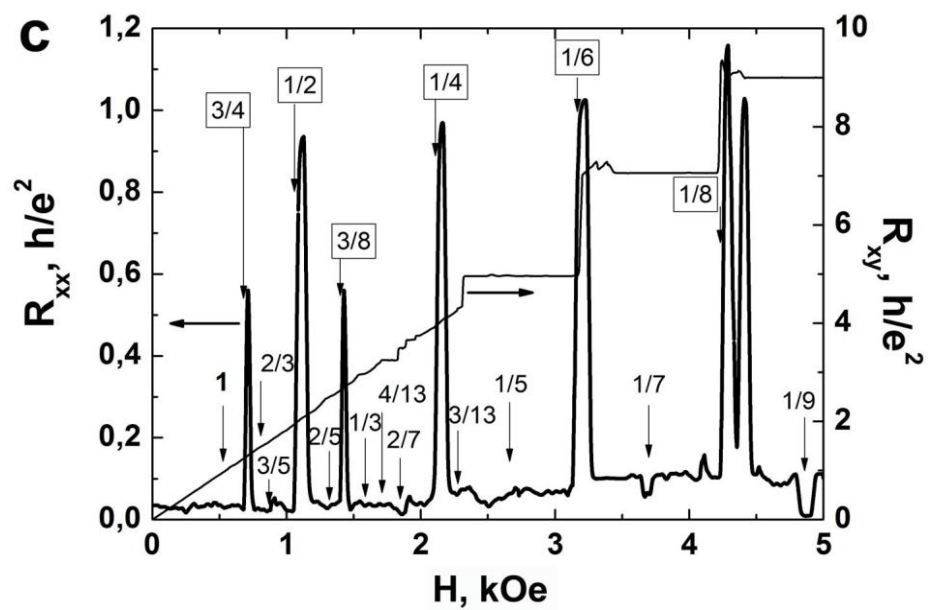

**Figure 5. Quantum Hall effect.**
The Hall resistance $R_{xy}=V_{xy}/I_{ds}$ and the magnetoresistance $R_{xx} = V_{xx}/I_{ds}$, $I_{ds}= 10$ nA, of a two-dimensional hole system, $p_{2D}=3\times10^{13}$m$^{-2}$, in silicon nanosandwich at the temperature of 77K vs magnetic field. (b) – Manifestation of the fractional quantum Hall effect near $v=1/2$, $p_{2D}=3\times10^{13}$m$^{-2}$. (c) - $R_{xx}$ and $R_{xy}$ vs magnetic field at the temperature of 77K, $I_{ds}= 10$ nA, for low sheet density of holes, $p_{2D}=1.3\times10^{13}$m$^{-2}$, that is revealed by varying the value of the top gate voltage (Fig. 1h), $V_{tg}=+150$ mV, see Ref. 11.

In addition to the aforesaid, the DX- and oxygen –related centers as well as the antisite donor-acceptor pairs reveal the negative-$U$ properties to confine effectively the edge channels in III-V compound low-dimensional structures[24-27]. However the preparation of dipolar configurations in this case appears to be accompanied by preliminary selective illumination at low temperatures.

Nevertheless, the question arises – how is possible to measure these quantum effects at high temperatures in weak magnetic fields? One of the reasons that are noted above is small effective mass, $10^{-4}m_0$, which is considered within the concept of the squeezed silicon arising owing to the negative-$U$ impurity stripes in edge channels (Fig. 6)[11]. Other reason is caused by very effective self-cooling inside negative-$U$ dipole boron strata under the conditions of the drain-source current.

In order to interpret the experimental data presented above, we propose the following thermodynamic model of the self-cooling process in the silicon nanosandwich. In the frameworks of this model, the whole thermodynamic process consists of multiple self-cooling cycles, each of which includes the following two stages: A) the adiabatic electric depolarization process of the spontaneously polarized stripe that is due to the presence of holes in the edge channel, and B) the fast spontaneous isothermal polarization of the negative-$U$ boron dipoles in the stripe with transition back into the ferroelectric state.

The model presented is based on 1) the energy conservation law in the electric depolarization process, and 2) the expression for the free energy, $f$, and the internal energy, $u$, of the ferroelectric state related to the thermodynamic work reserved in the edge channel of the silicon nanosandwich,

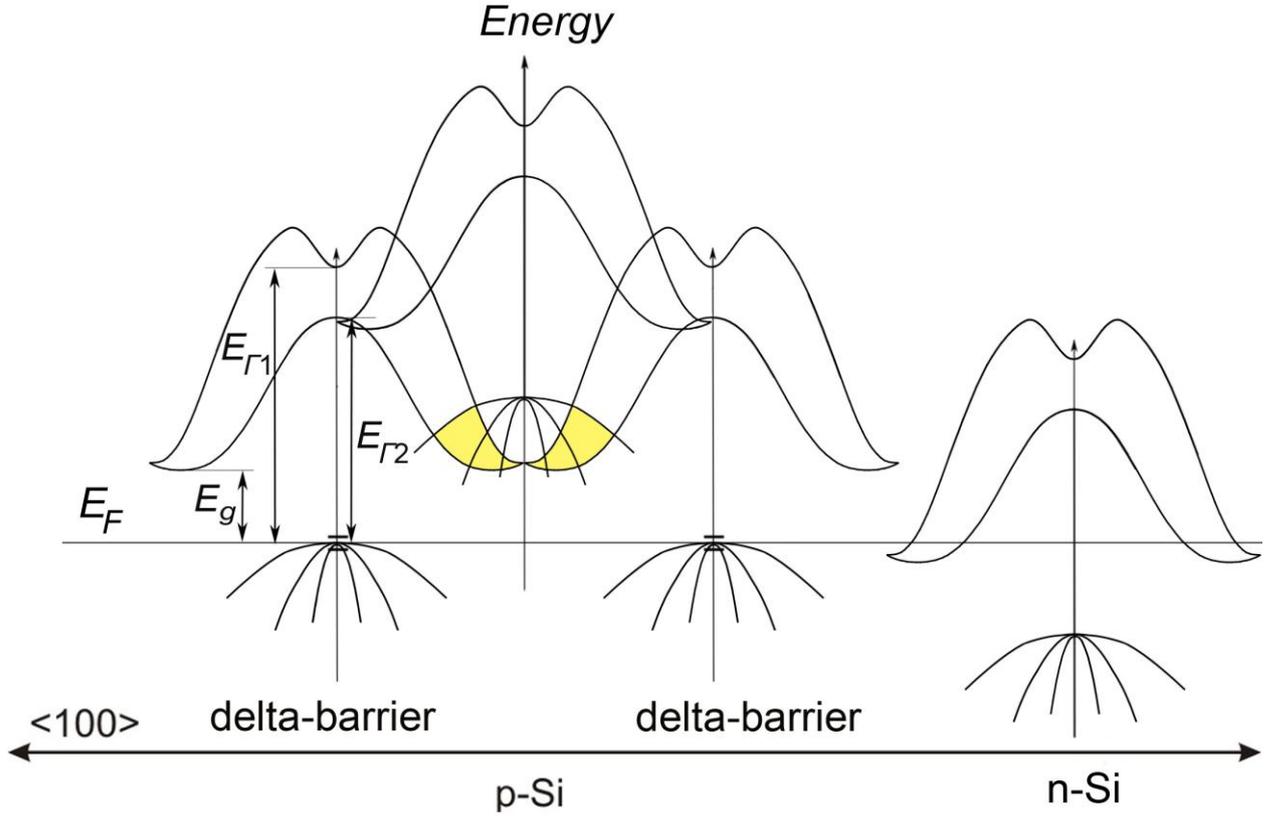

**Figure 6. One-electron band scheme of the silicon nanosandwich.**
The one-electron band scheme of the *p*-type Si-QW confined by the delta barriers heavily doped with boron on the *n*-type Si (100) surface. The delta-barriers are of wide-gap as a result of the formation of the negative-*U* dipole boron centers (see also Fig. 1d). The negative-*U* impurity stripes are of importance to result in the squeeze of the *p*-type Si-QW between the delta-barriers. In the frameworks of the squeezed silicon, the anti-crossing between the conduction band of the delta barriers and the valence band of the *p*-type Si-QW becomes to be actual thereby giving rise to the small effective mass of holes[11].

$$\delta Q = T\delta s = \delta u + \int E \delta P \, dV \quad (1)$$

$$u = -T^2 \frac{\partial}{\partial T}\left(\frac{f}{T}\right) = \int T^2 \frac{\partial}{\partial T}\left(\frac{EP}{T}\right) dV \quad (2)$$

Here *T* and *s* are the local effective temperature and the entropy of the nanosystem under consideration.

Combining the equations 1 and 2, we obtain the fundamental relationship describing a single thermodynamic self-cooling cycle of the negative-*U* stripe that is induced by the drain-source current in the edge channel:

$$c_P dT = -T\left(\frac{\partial E}{\partial T}\right)_P dP$$
$\Delta s = 0$ (3)

Here $E$, $P$ and $c_p$ are related to the local electric field, spontaneous dielectric polarization and the local lattice thermocapacity of the negative-$U$ stripe. Integration reveals a significant cooling effect in the edge channel of the silicon nanosandwich.

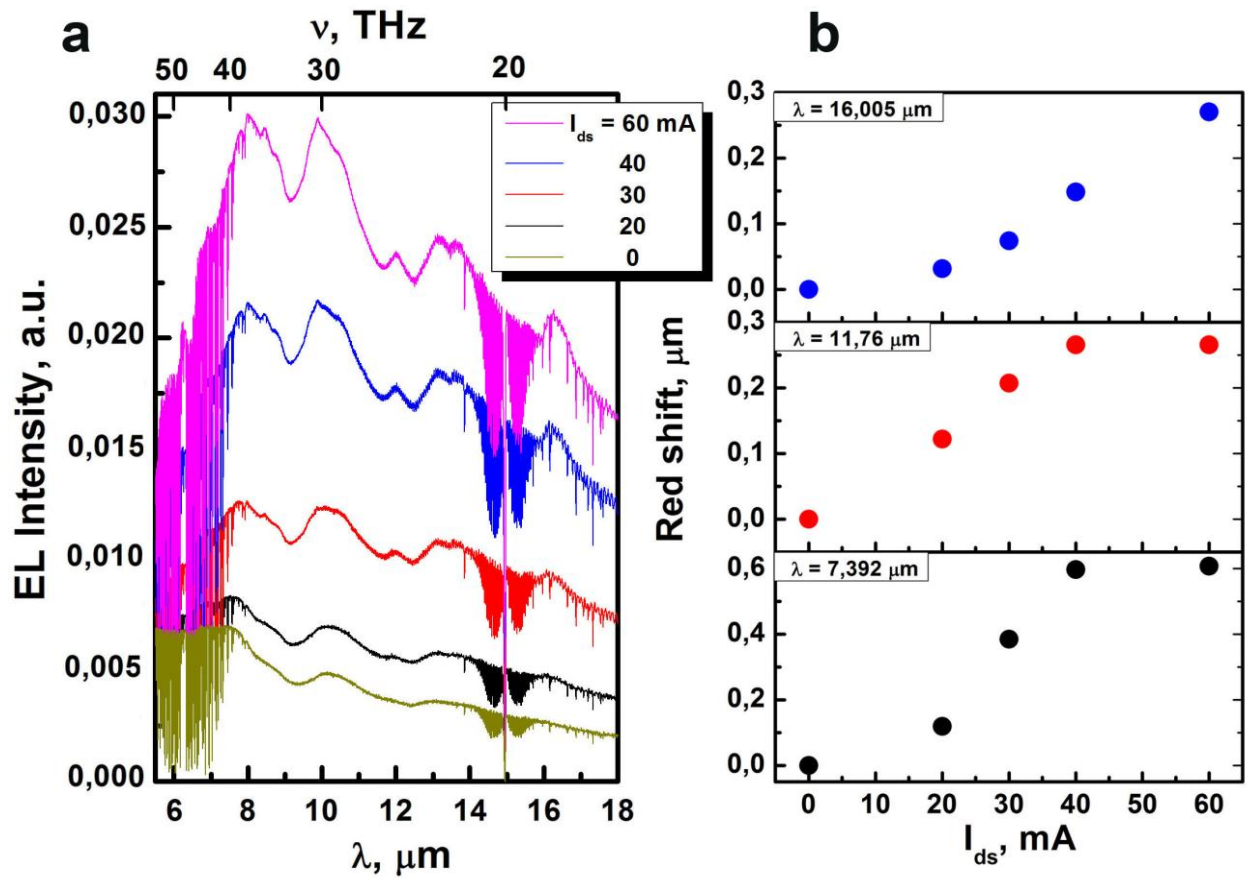

**Figure 7. The red shift of the radiation from the edge channels of silicon nanosandwich.**
(a) and (b) - The cooler effect is identified by the red shift of the radiation caused by the dipole boron centers from the edge channels of the silicon nanosandwich device (see Fig. 1h) that is induced by the stabilized drain-source current in different spectral range; T=300 K.

Thus, this model allows our interpretation of high temperature quantum kinetic effects in silicon nanosandwiches confined by the negative-$U$ ferroelectric or superconductor barriers. Specifically, the self-cooling effect of the edge channels can be derived as a red shift in the electroluminescence spectra that is enhanced by increasing the drain-source current, see Fig. 7.

## SUMMARY

Negative-$U$ impurity stripes containing single holes appear to reduce the electron-electron interaction in the edge channels of the silicon nanosandwiches that allow the observations of quantum kinetic effects at high temperatures up to the room temperature.

The edge channels of semiconductor quantum wells confined by the subsequence of negative-$U$ impurity stripes containing single holes give rise to the creation of the composite bosons and fermions in weak magnetic fields by the step-by-step capture of magnetic flux quanta.

The strong diamagnetism of the impurity stripes that consist of the negative $U$ dipole boron centers surrounding the Si-QW edge channels allowed the high temperature observation of the de Haas – van Alphen oscillations as well as integer and the fractional quantum Hall effect.

Finally, it should be noted that the step-by-step capture of magnetic flux quanta on all negative $U$ impurity stripes in the edge channels is determined by the balance between the numbers of composite bosons and fermions that is very interesting for models of quantum computing.

## ACKNOWLEDGEMENTS

The work was supported by the programme «5-100-2020», project 6.1.1 of SPSPU (2014); project 1963 of SPbGPU (2014); the programme of fundamental studies of the Presidium of the Russian Academy of Sciences "Actual problems of low temperature physics" (grant 10.4); project 10.17 "Interatomic and molecular interactions in gases and condensed matter".

# REFERENCES

1. Ezawa, Zyun F. *Quantum Hall Effects: Recent Theoretical and Experimental Developments* (World Scientific, Singapore 2013).

2. Eisenstein, J. P. et al. Density of States and de Haas —van Alphen Effect in Two-Dimensional Electron Systems. *Phys. Rev. Lett*. **55**, 875-878 (1985).

3. Landwehr, G. et al.Quantum transport in n-type and p-type modulation-doped mercury telluride quantum well. *Physica E* **6,** 713-719 (2000).

4. Novoselov, K.S. et al. Two-dimensional gas of massless Dirac fermions in graphene, *Nature* **438**, 197-200 (2005).

5. Novoselov, K. S. et al. Room-Temperature Quantum Hall Effect in Graphene. *Science* **315** (5817), 1379 (2007).

6. Geim, A.K. & Novoselov, K. S. The rise of grapheme. *Nature Materials* **6**, 183-191 (2007).

7. Hasan, M.Z. & Kane, C.L. Colloquium: Topological insulators. *Rev. Mod. Phys*. **82**, 3045-3067 (2010).

8. Klinovaja, J., Stano, P., Yazdani, A. & Loss, D. Topological Superconductivity and Majorana Fermions in RKKY Systems. *Phys. Rev. Lett.* **111**, 186805-5 (2013).

9. Zyuzin, A.A. & Loss, D. RKKY interaction on surfaces of topological insulators with superconducting proximity effect. *Phys. Rev.B* **90**, 125443-5 (2014).

10. Bagraev, N. et al. Self-assembled impurity superlattices and microcavities in silicon. *Def. Dif. Forum* **194-199**, 673-678 (2001).

11. Bagraev, N.T., Galkin, N.G., Gehlhoff, W., Klyachkin, L.E. & Malyarenko, A.M. Phase and amplitude response of "0.7 feature" caused by holes in silicon one-dimensional wires and rings. *J. Phys.: Condens. Matter*. **20**, p.164202-12 (2008).

12. Gehlhoff, W., Bagraev, N.T. & Klyachkin, L.E. Cyclotron resonance in heavily doped silicon quantum wells. *Sol. St. Phenomena* **47-48**, 589-594 (1995).

13. Bagraev, N.T., Klyachkin, L.E., Kudryavtsev, A.A., Malyarenko, A.M., & Romanov, V.V. *Superconductor* Ch. 4 (SCIYO, Croatia, 2010).

14. Bagraev, N.T., Ivanov, V.K., Klyachkin, L.E. & Shelykh, I.A. Spin depolarization in quantum wires polarized spontaneously in a zero magnetic field. *Phys. Rev. B* **70**, 155315-9 (2004).

15. Bagraev, N.T., et al. Shubnikov–de_Haas and de_Haas–van_Alphen Oscillations in Silicon Nanostructures. *Semiconductors* **45**, 1447–1452 (2011).

16. Bagraev, N.T. et al. EDESR and ODMR of Impurity Centers in Nanostructures Inserted in Silicon Microcavities, *Appl. Magn. Reson.* **39,** 113–135, (2010).

17. Laughlin, R. B. Quantized Hall conductivity in two dimensions. *Phys. Rev. B* **23**, 5632-5633 (1981).

18. Laughlin, R. B. Quantized motion of three two-dimensional electrons in a strong magnetic field. *Phys. Rev. B* **27**, 3383-3389 (1983).

19. Halperin, B.I. Theory of the Quantized Hall Conductance. *Helvetica Physica Acta* **56,** 75-102 (1983).

20. Dolgopolov, V.T. Integer quantum Hall effect and related phenomena. *Phys. Usp*. **57** 105–127 (2014).

21. Devyatov, E.V. Edge states in the regimes of integer and fractional quantum Hall effects. *Phys. Usp.* **50**, 197–218 (2007).

22. von Klitzing, K., Dorda, G. & Pepper, M. New method for high-accuracy determination of the fine-structure constant based on quantized Hall resistance. Phys. Rev. Lett. 45, 494-497 (1980).

23. Willett, R. et al. Observation of an Even-Denominator Quantum Number in the Fractional Quantum Hall Effect. *Phys. Rev. Lett*. **59**, 1776-1779 (1987).

24. Zazoui, M. Feng, S. L. & Bourgoin, J. C. Nature of the *DX* center in $Ga_{1-x}Al_xAs$, *Phys. Rev. B* **44**, 10898-10900 (1991).

25. Peale, R. E. Mochizuki, Sun, Y.H. & Watkins, G. D. Magnetic circular dichroism of the *DX* center in $Al_{0.35}Ga_{0.65}As$:Te, *Phys. Rev. B* **45**, 5933-5943 (1992).

26. Alt, H. Ch. Experimental evidence for a negative-*U* center in gallium arsenide related to oxygen, *Phys. Rev. Lett.* **65**, 3421-3425 (1990).

27. Bagraev N.T., The EL2 center In GaAs: Symmetry and Metastability, *J. Phys. (France) I*, **1**, 1511-1527 (1991).